# Easy JavaScript Simulation (EJSS) Data Analytics for Singapore

Loo Kang WEE[1], Darren TAN[1], Félix Jesús Garcia CLEMENTE[2], Francisco ESQUEMBRE[3]
[1] Ministry of Education, Singapore
[2] Computer Engineering and Technology Department, University of Murcia, Spain
[3] Department of Mathematics, University of Murcia, Spain
Lawrence_WEE@moe.gov.sg , Darren_TAN@moe.gov.sg , fgarcia@um.es ,
fem@um.es
Corresponding author's e-mail: Lawrence_WEE@moe.gov.sg

**Abstract.** We have integrated Easy JavaScript Simulation (EJSS) Data Analytics into the national Learning Management System for Singapore schools, known as the Singapore Student Learning Space (SLS). EJSS Data Analytics enhances the teaching and learning experience for educators and students by enabling educators to monitor and evaluate students' interactions with interactive computer simulations. The data analytics and visualisation capabilities are delivered using the Moodle platform and version 1.3 of the specifications for Learning Tools Interoperability (LTI). In this paper, we showcase the potential for EJSS Data Analytics to identify students' learning difficulties and misconceptions. Four examples of EJSS Data Analytics applications are provided to illustrate insights on aspects that include understanding a student's sequential actions leading to specific task outcomes, the frequency of task attempts by each student, and the ratio of students achieving correct versus incorrect task completions. We identify five key considerations for designing the EJSS teacher dashboard. These considerations relate to Student Thought Process, Student Behaviour, Student Engagement, Student Choice, and Teacher Feedback. These five facets provide a framework for aligning our design efforts with the needs of students and teachers, also drawing upon research in data analytics for education.

## 1. Introduction

*1.1. Overview*
The Easy JavaScript Simulation (EJSS) authoring toolkit, formerly known as the Easy Java Simulation (EJS) authoring toolkit, has evolved significantly since its inception in 2000. From its origins as a toolkit for educators to produce java applets, it now supports HTML5 [1] remote web-based-control laboratories [2] mobile apps [3] sensor-aware apps [4] and data analytics [5].

This paper presents our work on integrating the EJSS App into a Learning Management Systems (LMS) that support Learning Tools Interoperability (LTI) version 1.3. This integration will enable educators to harness the power of EJSS Data Analytics to better understand and evaluate students' actions within simulations.

*1.2. Problem and Context*
Understanding and evaluating students' actions within simulations is a long-standing challenge for educators. Identifying specific misconceptions or errors made by students within these simulations remains elusive. Typically, instructors are required to invest valuable time in instructing students to replicate these "mistakes" to better discern the necessary interventions. The Singapore Student Learning Space (SLS) is a key initiative by the Ministry of Education (MOE) to transform the learning experiences of students through the purposeful use of technology for learning modes like self-directed and collaborative learning. The EJSS App works within SLS as a "Class Assignment" available in most Learning Management System (LMS).

*1.3. Solution*

The integration of the EJSS App with the SLS platform empowers teachers with data and visualisations on individual students' interactions with simulations. This valuable information enables educators to better assess students' comprehension, potentially refining their instructional approaches. Notably, the EJSS App is unique within SLS in providing student interaction data for computer simulations. This data may encompass a variety of insights, including a student's sequential actions leading to specific task outcomes, the frequency of task attempts, or the ratio of students achieving correct versus incorrect task completions. It is currently not possible to collect and analyse such data within SLS when using alternative interactive platforms like PhET or GeoGebra.

## 2. Examples of EJSS Data Analytics

Each simulation that is compatible with the EJSS App is associated with a teacher dashboard, designed to allow the teacher to visualise their students' interactions in the simulation.

*2.1. Vectors*

Figure 1 shows the view of an EJSS where student interact with the simulation on the topic of vectors, given the horizontal and vertical components of a vector, to calculate either angle $\vartheta$ or $\phi$ with horizontal or vertical respectively. Figure 2 shows the teacher dashboard suggests that the annotated student is likely mistaking the angle $\phi$ with its complementary angle ($90° - \phi$). By looking at similar randomly generated questions such as Q1 and Q3, the teacher can help the student recognise the pattern of errors.

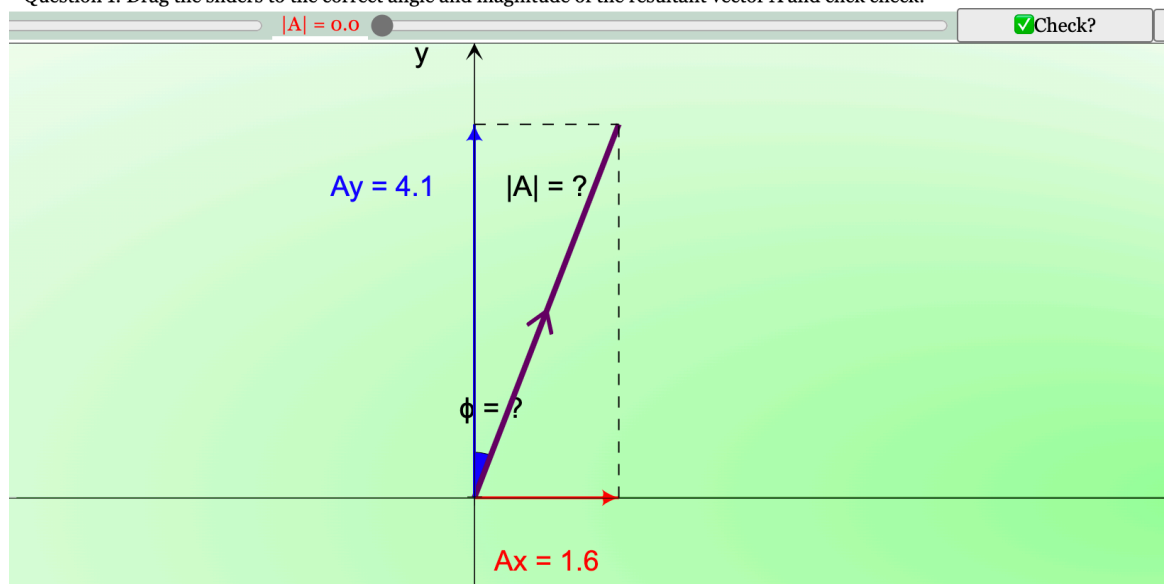

**Figure 1.** View of EJSS Vectors Question 1: vector *A* horizontal component *Ax* = 1.6, vertical component *Ay* = 4.1, find |*A*| and $\phi$ (left).

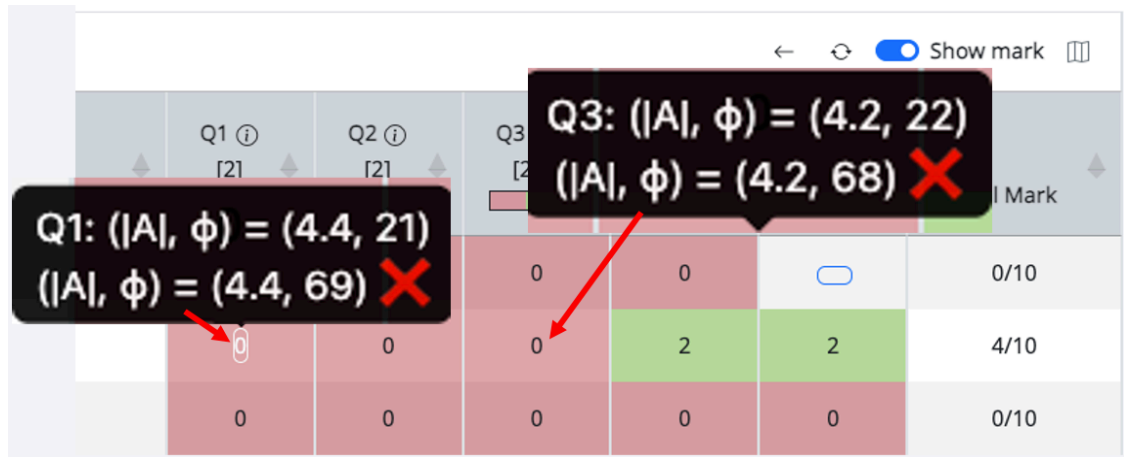

**Figure 2.** The teacher dashboard view (right), showing Question 1 where the student's answer for $|A| = \sqrt{1.6^2 + 4.1^2} = 4.4$ is correct but $\phi = \tan^{-1}\left(\frac{1.6}{4.1}\right) = 21°$ is incorrect. The pattern of data in both Q1 and Q3 suggests a consistent misconception about the angle $\phi$ indicated as $(90° - \phi)$.

*2.2. Projectile Motion (Frog Game)*

Figure 3 shows the EJSS where students interact with the projectile frog game by changing the angle of launch $\vartheta$, to complete stage 3 of the game. The desired student learning method is to use the two-component projectile motion equation (1) to solve for $\vartheta$ and key in into the corresponding field to check for the correctness of their answer. The solution lies in solving Equation (1), where $x = 2$, $y = 0$, $u = 10$, $g = 9.81$, simplifying to get $t = 0$ or $\frac{10\sin\vartheta}{g}$. Further solving to get $\frac{2}{100}g = \sin 2\vartheta$, therefore, $\vartheta = 5.7°$ or $84.3°$

$$\begin{pmatrix} x \\ y \end{pmatrix} = \begin{pmatrix} u\cos\vartheta \\ u\sin\vartheta \end{pmatrix} t + \frac{1}{2}\begin{pmatrix} 0 \\ -g \end{pmatrix} t^2 \tag{1}$$

Without the teacher dashboard (figure 4) and without asking each student to report how they found the angle, this evidence-based insight can only be observed by watching the student's screen.

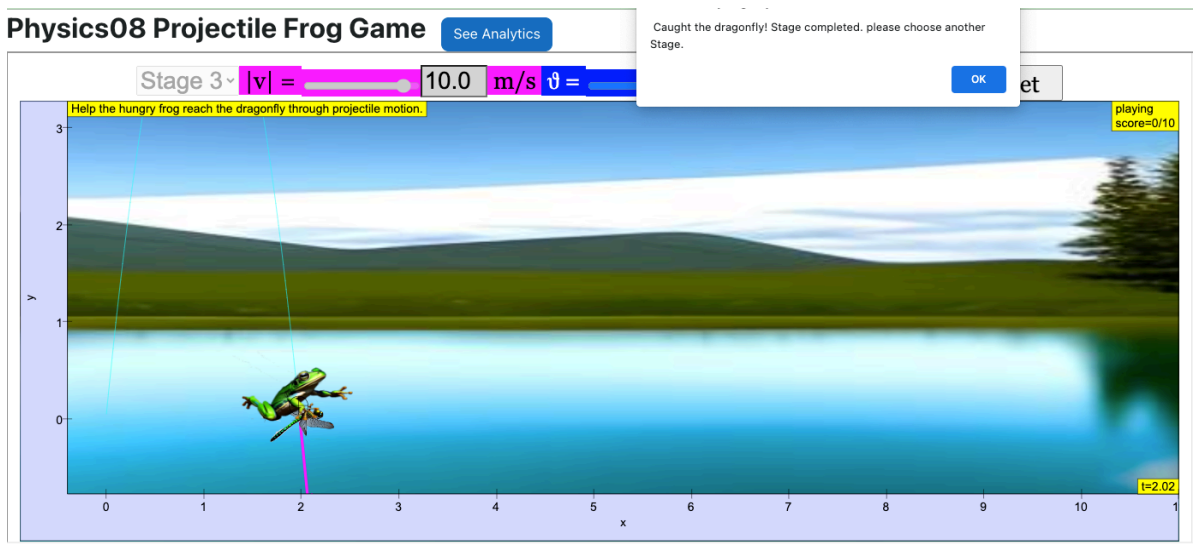

**Figure 3.** View of EJSS Projectile Motion (Frog Game) Stage 3: Frog is launched at position (0,0) with a fixed speed $|v| = 10$ m/s, dragonfly is at position (0,2) and student is to find $\vartheta$.

**Figure 4.** The data analytics suggests the student did not attempt to solve analytically but rather used a trial-and-error approach.

*2.3. Eight-point Compass*
In this Primary Mathematics (figure 5) example students are tasked to click and drop the respective animals into the correct position referencing the origin (0,0). The data (figure 6) suggests that student "Student100" has difficulty with East and West when combined with the North and South direction. The student "slscgbot" may have difficulty estimating numbers larger than 5.

**Figure 5.** View of EJSS Eight-Point Compass: Student are tasked to click and drop the different animals into the correct position from origin (0,0).

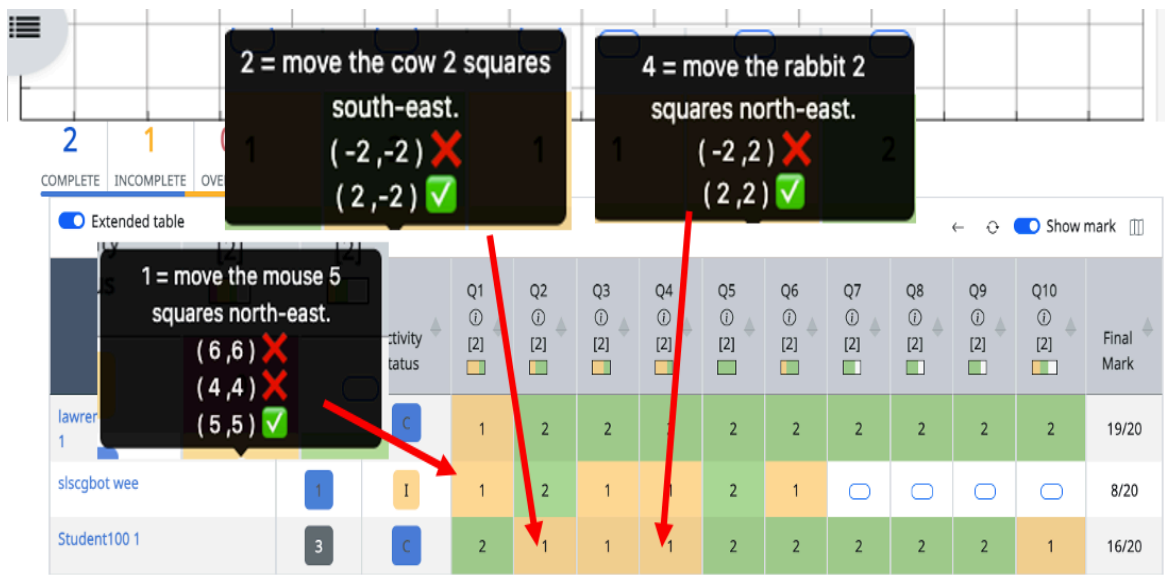

**Figure 6.** Looking at "Student100" attempt on Q2 and Q4, these 2 data suggest student has performance difficulties for 2 (North-South with East-West) directions coordinates. Next, "slscgbot" on Q1, the task was to move the mouse 5 squares north-east, the moves are (6,6) and (4,4) for first and second attempts, suggesting performance difficulties to get the correct (5,5).

### 2.4. 21st Century Competencies (21CC) Navigator

In this 21st Century Competencies (21CC) Navigator Future Self Quiz example (figure 7), the EJSS data analytics capability (figure 8) is used to capture survey information accurately and automatically.

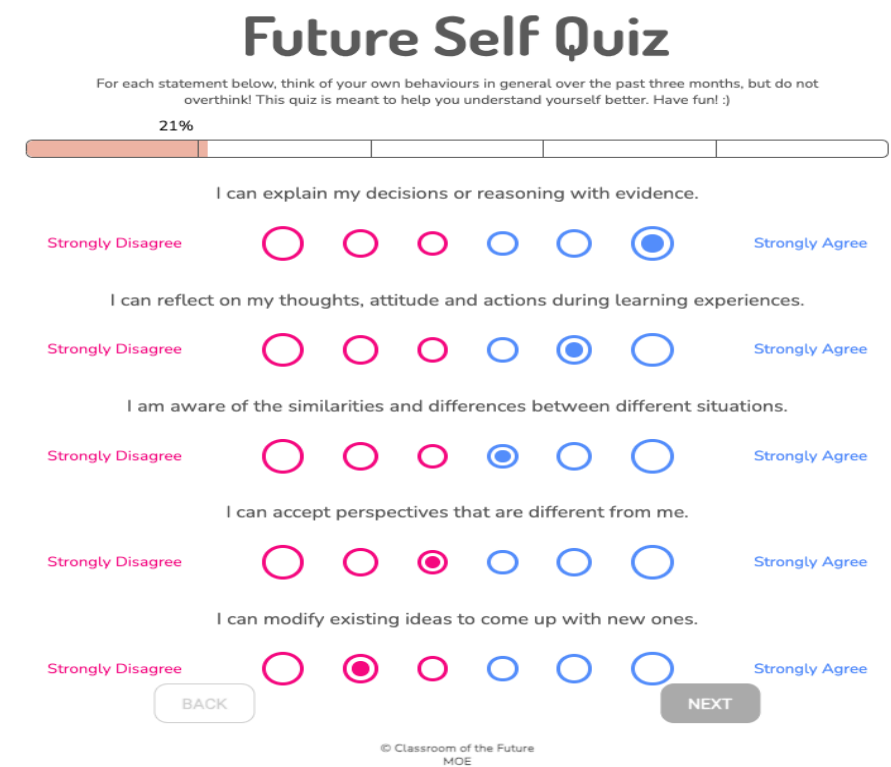

**Figure 7.** View of EJSS 21CC Navigator Future Self Quiz survey: Student are tasked to click on all 24 survey questions.

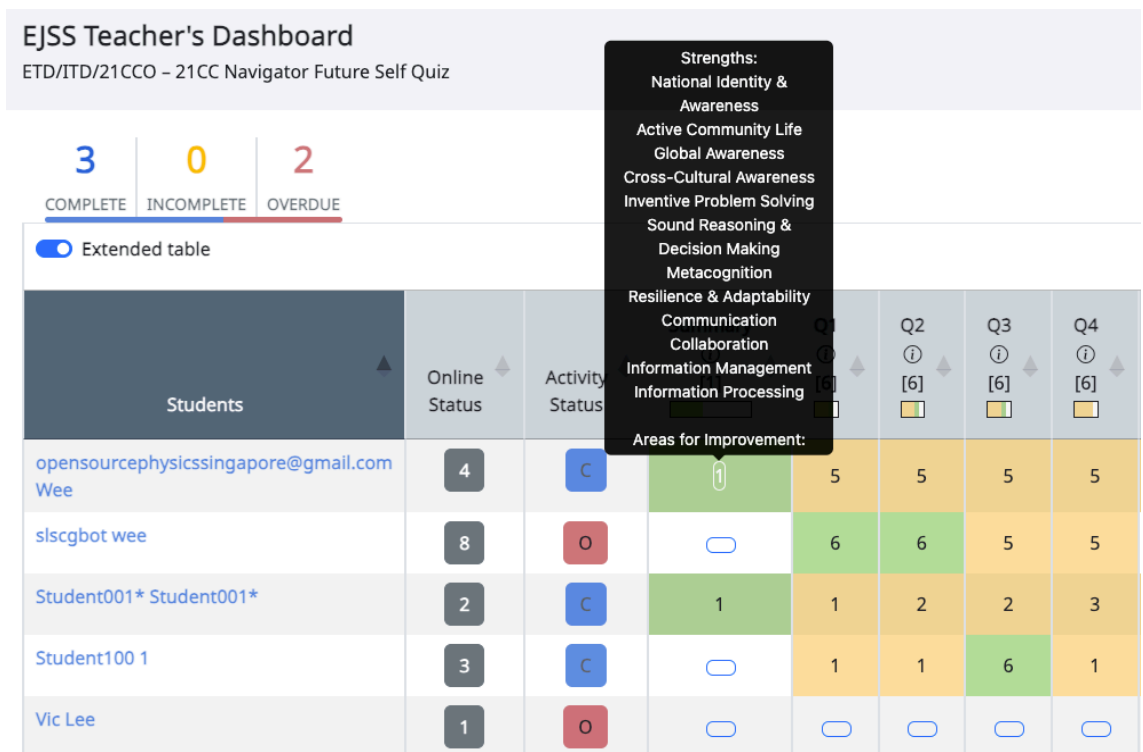

**Figure 8. EJSS Teacher's Dashboard, the teacher can see each student's Areas for Improvement without need for students to manually copy and paste their Strengths and Areas for Improvement into SLS textboxes for data collection.**

Assigning 1 mark to "Strongly Disagree" and 6 marks to "Strongly Agree" gives a total of 24 marks for Q1 to Q4. Based on a 58% cut-off, marks below 13.92 are classified as "Areas for Improvement" while marks above that are "Strengths". This ability to calculate and automatically score and assign as Strengths or Areas for Improvement power up the functionality of SLS, ensuring appropriate intervention.

**3. Design Considerations for the EJSS Teacher Dashboard**

We discuss design considerations for the EJSS Teacher Dashboard, linking our simulation design scholarly work with learning analytics literature [6,7].

*3.1. Student Thought Process*

The tooltip (figure 9) shows the sequence of stages leading to both accurate and misguided responses. This chronological dossier of the student's interactions in the simulation furnishes educators with insights into students' cognitive process.

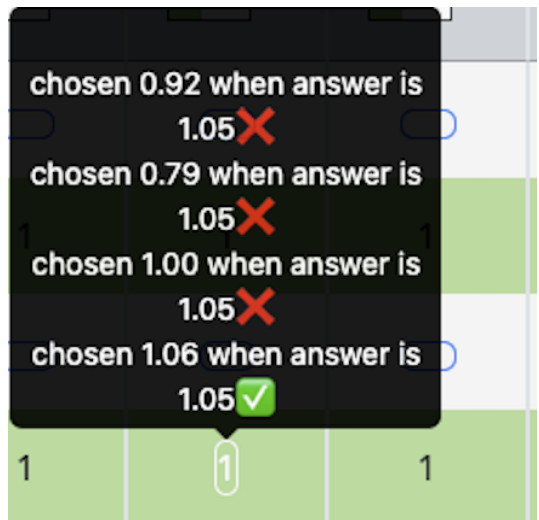

**Figure 9.** Sequence of stages in chronological order, student chose 0.92, followed by 0.79, followed by 1.00 and finally 1.06 which is close enough to the correct answer of 1.05.

*3.2. Student Behaviour*

We devised a metric of active and inactive time (figure 10) to provide a glimpse into students' work ethic. The teacher can use the data to praise effort and promote positive classroom culture.

| Students | Online Status | Activity Status |
|---|---|---|
| AFT | 6 | C |
|  | 2 | I |
|  | 3 | C |
|  | 3 | I |

**Figure 10.** The numbers 6, 2, 3, 3 are the number of attempts by the students. The higher the number, it is an indication of the level of diligence exhibited as it logged number of attempts on the EJSS.

*3.3. Student Engagement*

We established a system for retaining recent re-attempts and introduced a gamified marking scheme (figure 11). Our observations indicated that this mechanism increased students' motivation to revisit and enhance their performance in these tasks.

**Figure 11.** For example, the third student Q4 is 1 mark (in orange), this lower mark may increase the motivation of the students to revisit the same Q4 and try to get the answer correct, after 1st re-attempt which will award the student 2 marks (in green). Similarly, the 0 marks (in red) will have a similar effect on students' motivation to re-attempt for a higher final mark. Also note that student can chose unrestricted by any predetermined sequence, Q1 and Q3 are not attempted, while student can attempt Q4, Q2 and Q0 which we believe increases student motivation through better user experience and greater choice.

*3.4. Student Choice*

We engineered a feature that empowers students to navigate freely among questions, unrestricted by any predetermined sequence (figure 11). We discovered that this flexibility enhanced students' overall learning experience.

*3.5. Teacher Feedback*

We introduced a visual indicator (figure 12) that shows the percentage of students who provided incorrect answers, which was a commonly requested feature. This indicator bar is intuitive, obviating the necessity for teachers to scroll through and manually tally instances of incorrect responses, saving valuable classroom time and precious cognitive effort.

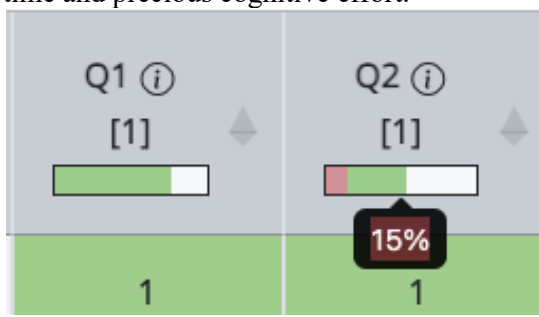

**Figure 12.** We implemented a visual bar showing 15% for example for the percentage of students who were awarded 0 mark to support teachers' feedback for a more user-friendly display to quickly identify which questions to discuss in class.

## 4. Integration Overview

*4.1. Overview*

Once integrated within a Learning Management System (LMS) compliant with LTI 1.3 standards (figure 13), educators can seamlessly assign courses with EJSS simulations to student groups. As students access the course that includes EJSS simulations, the Moodle server will facilitate automatic authentication for Moodle-EJSS usage. For optimal results, it is advisable to employ Moodle version 4.02 as the LTI provider, effectively delivering EJSS content to any LMS such as Singapore Learning Space (SLS) as the LTI consumer.

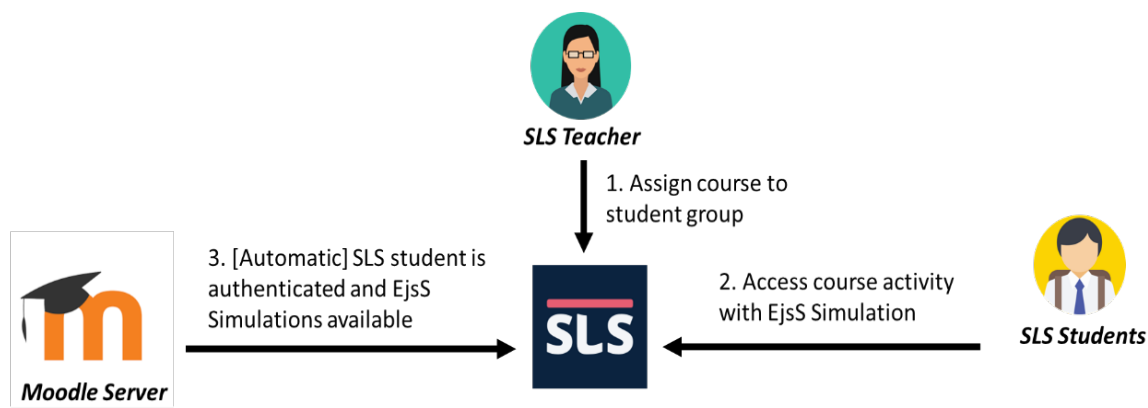

**Figure 13.** Overview of a SLS Teacher assign a course with EJSS to a student group that is automatically authenticated with our Moodle Server and the EJSS is available to SLS students.

*4.2. Pre-requisites*

To enjoy these data analytics, certain pre-requisites must be fulfilled:
- For EJS Simulations created between 2000 and 2013: The outdated Java-based *.jar files should be transformed into modern JavaScript EJSS format *.zip files.
- For EJSS Simulations generated from 2013 to the present: These simulations need to be regenerated using the javascript_ejs_6.02_beta_201228.zip toolkit.
- Additional questionLib custom function will need to be added.
- Incorporation of the assessment.json file is required, defining grading criteria (0, 1, or 2) and preserving interaction history per question. A Python file is available for generating this file.
- A Moodle 4.02 server, for example https://iwant2study.org/moodle402/ installed with the EJSS learning analytics plugin https://gitlab.com/ejsS/addon-projects/moodle-plugin.
- Configure an LTI App consumer within your own LMS such as Singapore Learning Space (SLS) and an LTI App provider within the Moodle-EJSS system. This step can be skipped if the LMS is Moodle and the plugin works inside without the need for LTI tool publishing.

## 5. Conclusion

The EJSS authoring toolkit, renowned for its openness and multiple awards, now features data analytics functionality, seamlessly integrated for utilization within LTI 1.3-compliant LMS such as Singapore's SLS. Four examples of EJSS Data Analytics were discussed, with insights encompassing a diverse range of aspects, including a student's sequential actions leading to specific task outcomes, the frequency of task attempts, and the ratio of students achieving correct versus incorrect task completions.

Five key considerations were identified in relation to designing the EJSS teacher dashboard. These considerations revolve around understanding and addressing Student Thought Process, Student

Behaviour, Student Engagement, Student Choice, and Teacher Feedback. These facets serve as a framework, aligning our design efforts with the unique needs of both students and teachers while drawing upon research in data analytics. Lastly, we explain integration pre-requisites, knowledge and tools they need to leverage EJSS data analytics effectively in their educational context. We hope to provide the necessary know-how to improve EJSS data analytics usage across other educational systems.

**Acknowledgments**
This EJSS App is funded by the Singapore Ministry of Education (MOE) Senior Specialist Research - Development Fund (SSTRF).